\documentclass[preprint, double-spaced]{revtex4}
\usepackage{graphicx}
\usepackage{amsmath,bbm}
\usepackage{amssymb,bm}
\usepackage{array}

\begin{document}
\title{Color screening scenario for quarkonia suppression in a quasiparticle model compared with data obtained from experiments at the CERN SPS, BNL RHIC, and CERN LHC}
\author{P.~K.~Srivastava\footnote{$prasu111@gmail.com$}$^1$}
\author{M. Mishra$^2$}
\author{C.~P.~Singh$^1$}

\affiliation{$^1$Department of Physics, Banaras Hindu University, 
Varanasi 221005, INDIA}
\affiliation{$^2$Birla Institute of Technology and Science, Pilani - 333031, INDIA}

\begin{abstract}

We present a modified colour screening model for $J/\psi$ suppression in the Quark-Gluon Plasma (QGP) using quasi-particle model (QPM) as equation of state (EOS). Other theoretical ingredients incorporated in the model are feed-down from higher resonances namely, $\chi_c$, and $\psi^{'}$, dilated formation time for quarkonia and viscous effects of the QGP medium. Assuming further that the QGP is expanding with Bjorken's hydrodynamical expansion, the present model is used to analyze the centrality dependence of the $J/\psi$ suppression in mid-rapidity region and compare it with the data obtained from SPS, RHIC and LHC experiments. We find that the centrality dependence of the data for the survival probability at all energies is well reproduced by our model. We further compare our model predictions with the results obtained from the bag model EOS for QGP which has usually been used earlier in all such calculations.
\\

\end{abstract}

\maketitle 
\section{Introduction}
\noindent
Ultra-relativistic collisions of heavy nuclei are believed to produce nuclear matter at extreme pressure and energy density. Quantum Chromodynamics (QCD) predicts that a deconfined state of partonic matter which is referred as the quark-gluon plasma (QGP), will be produced in such circumstances. One of the most striking signatures of QGP formation is the suppression of heavy quarkonia states such as the charmonia ($J/\psi,~\psi^{'},~\chi_{c}$ etc) in the deconfined phase. Till mid-2000, Debye screening was thought to be the only possible mechanism for the anomalous quarkonia suppression in QGP but recent work by Laine and collaborators have shown that the effective potential between heavy quarks acquires an imaginary part as well~\cite{ima1,ima2,ima3}. This imaginary part generates the collisional dissociation of bound states by the QGP~\cite{ima1,ima2,ima3}. However, in this paper, we emphasize the Debye screening as the dominant mechanism for the quarkonia suppression in the QGP medium. Charmonia suppressions in heavy-ion collisions have experimentally been studied, first at the CERN Super Proton Synchrotron (SPS) by NA50~\cite{na50} and NA60 experiment~\cite{na60} at $\sqrt{s_{NN}}=17.3-19.3$ GeV and then at Relativistic Heavy Ion Collider (RHIC) by the PHENIX experiment at  $\sqrt{s_{NN}}=200,~39$ and $62.4$ GeV ~\cite{ada}, and now at Large Hadron Collider (LHC) by the CMS and ALICE experiments at $\sqrt{s_{NN}}=2.76$ TeV ~\cite{hepe,hepe1}. However, experimental results involve more puzzling features which defy explanations on the basis of colour-screening in the QGP alone. For example, we notice significantly less suppression at the mid-rapidity than at forward rapidity ($1.2< |y|< 2.2$) while the medium in the central part of the collision is the most dense and intutively one should expect an opposite behaviour. Another interesting experimental observation is similar $J/\psi$ suppression at SPS and RHIC energies for the same number of participants~\cite{ada}. Many theoretical efforts have since been made which explore the effects of cold nulear matter (CNM) as well as possible coalescence of originally uncorrelated $c$ , $\bar c$ quark pairs. However, a complete theoretical interpretation for $J/\psi$ suppression is still lacking.

  Many theoretical papers have analyzed the heavy-ion experimental data regarding nuclear modification factor for heavy quarkonia. For example, $Pb+Pb$ data at SPS are explained by variety of models~\cite{blaiz,blaiz1,capel,akch,akch1,akch2,akch3,grand} with or without deconfinement phase transition scenario. Recently Yunpeng et al.,~\cite{yun} analyzed the $J/\psi$ suppression data using a transport model calculation involving suppression as well as the regeneration of $J/\psi$ due to coalescence of $c$ and $\bar c$ pair produced abundantly in the initial stage of collisions at RHIC and LHC energies. Zhen Qu et al.,~\cite{zhen} used a similar approach to explain the forward rapidity data for $J/\psi$ at RHIC energy. Sharma and Vitev~\cite{rishi} calculated the yields of quarkonia at RHIC and LHC as a function of transverse momentum based on heavy quark effective theory in which both colour singlet and colour octet contribituions along with feed down effects from excited states were suitably incorporated. Their model provides good description of the LHC data for central as well as peripheral collisions. But none of the available models could provide a satisfactory and consistent explanation for the suppression at all energies simultaneously. We have recently modified~\cite{mmish} the colour-screening model of Chu and Matsui~\cite{chu} by parametrizing pressure instead of energy density since pressure density becomes almost zero at the deconfinment phase transition point. We have further shown that the variation of survival probability of $J/\psi$ with respect to participant number obtained from this model compares well with the experimental data (CNM normalized) at RHIC~\cite{mmish}. In the present paper, we reinvestigate the colour-screening model by employing a more realistic thermodynamically-consistent quasi-particle prescription for the equation of state (EOS) of QGP medium. This highlights a major difference between our present approach and the model of Chu, Matsui~\cite{chu} or Mishra et al~\cite{mmish} because bag model is often considered as a crude EOS for QGP. We have recently demonstrated that various thermodynamical as well as transport quantities obtained from the present quasiparticle model (QPM) compare well with the most recent available lattice results~\cite{pks, pks1}. In addition, we incorporate here viscous hydrodynamics in our formulation with a constraint that the ratio $\eta/s$ varies very slowly~\cite{pks1} with the temperature in the QGP medium. This assumption helps us in taking $\eta$ to be almost independent of the temperature. Feed down from higher resonances have also been incorporated in the model. 

  We use our formulation to explain the recent LHC experimental data on $J/\psi$ suppression along with the RHIC and SPS data.  We surprisingly find that $J/\psi$ suppression at mid-rapidity is explained consistentely by our model at SPS, RHIC and LHC energies in QGP picture alone. Furthermore, we compare results obtained from quasiparticle model (QPM) with those from the bag model EOS on $J/\psi$ suppression and find that QPM EOS describes the data better and thus lends credibility to the present approach in explaining the $J/\psi$ suppression. 
\section{Formulation}
\subsection{Cooling law}
The main theme of the present formulation is essentially borrowed from our previous model ~\cite{mmish} where we have used the colour screening idea of Chu and Matsui~\cite{chu}. However, instead of using bag model for QGP, we now use quasiparticle model (QPM) as new EOS of QGP. We have assumed that the QGP medium formed during the collision, expands and cools according to the Bjorken's boost invariant longitudinal hydrodynamics in mid-rapidity region. Employing the conservation of energy-momentum tensor, the rate of the decrease of energy density $\epsilon $~\cite{teany} is given by
\begin{equation}
\frac{d\epsilon}{d\tau}=-\frac{\epsilon+p}{\tau}+\frac{4\eta}{3\tau^2},
\end{equation}
where $\eta$ is the shear viscosity of the QGP medium, $p$ is the pressure and $\tau$ represents the proper time. The energy density and pressure are computed by using QPM EOS~\cite{pks} for QGP. Using Eq.(1) and the thermodynamical identity $\epsilon=T\frac{dp}{dT}-p$, the cooling laws for energy density and pressure in the QPM model can be separately given as (see Appendix A for derivation) :
\begin{equation}
\epsilon=c_1+c_2\tau^{-q}+\frac{4\eta}{3c_s^2}\frac{1}{\tau},
\end{equation}

\begin{equation}
p=-c_1+c_2\frac{c_s^2}{\tau^q}+\frac{4\eta}{3\tau}\left(\frac{q}{c_s^2-1}\right)+c_3\tau^{-c_s^2},
\end{equation}
where $c_1$ , $c_2$ and $c_3$ are constants which can be determined by imposing the initial boundary conditions on energy density and pressure, $q=c_{s}^{2}+1$ with $c_{s}$ as the speed of sound in the medium. We take $\epsilon=\epsilon_0$ at $\tau=\tau_0$ (initial thermalization time) and also $\epsilon=0$ at $\tau=\tau^{'}$; where $\tau^{'}$ is the proper time. Consequently, the constants $c_1$ and $c_2$ are related as :
\begin{equation}
c_1=-c_2\tau '^{-q}-\frac{4\eta}{3c_s^2\tau^{'}}
\end{equation}
, where $\tau^{'}=\tau_0 A^{-\frac{3R}{R-1}}$, $A=T_0/T^{'}$ and $R$ is the Reynold's number for QGP. Further :
\begin{equation}
c_2=\frac{\epsilon_0-\frac{4\eta}{3c_s^2}\left(\frac{1}{\tau_0}-\frac{1}{\tau^{'}}\right)}{\tau_0^{-q}-\tau^{'-q}}.
\end{equation}
Using the other initial condition for $p=p_0$ at $\tau=\tau_0$, gives the value of $c_3$ as follows :

\begin{equation}
c_3=(p_0+c_1)\tau_0^{c_s^2}-c_2c_s^2\tau_0^{-1}-\frac{4\eta}{3}\left(\frac{q}{c_s^2-1}\right)\tau_0^{(c_s^2-1)}.
\end{equation}
\subsection{Pressure Profile}
Assuming that the pressure almost vanishes at the transition temperature $T=T_c$, we take a pressure profile function in the transverse plane with a transverse distance $r$ as~\cite{mmish} : 
\begin{equation}
p(t_i,r)=p(t_i,0)h(r); \quad h(r)=\left(1-\frac{r^2}{R_T^2}\right)^{\beta}\theta(R_T-r),
\end{equation} 
where the coefficient $p(t_i,0)$ is yet to be determined, $R_T$ denotes the radius of the cylinderical plasma and it is related to the transverse overlap area $A_T$ as determined by Glauber model $R_T=\sqrt{\frac{A_T}{\pi}}$~\cite{balver,adler}. The pressure is thus assumed to be maximum at the central axis but it vanishes at the edge $R_{T}$ where hadronization first begins. The exponent $\beta$ depends on the energy deposition mechanism and here we have taken $\beta=1.0$~\cite{mmish}; $\theta$ is the unit step-function. The factor $p(t_i,0)$ is related to the average initial pressure $<p>_i$ via
\begin{equation}
p(t_i,0)=(1+\beta)<p>_i.
\end{equation} 
The average pressure is determined by the centrality dependent initial average energy density $<\epsilon>_i$ which is further given by Bjorken's formula~\cite{adler,bjor} :
\begin{equation}
<\epsilon>_i=\frac{1}{A_T\tau_i} \frac{dE_T}{dy}. 
\end{equation}
Here $dE_T/dy$ is the transverse energy deposited per unit rapidity. We use the experimental value of $dE_{T}/d\eta^{'}$ where $\eta^{'}$ is the pseudorapidity and then multiply it by a corresponding Jacobian factor~\cite{adler,cms1} in order to obtain $dE_{T}/dy$ for a given number of participants ($N_{part}$) at a particular center-of-mass energy ($\sqrt{s_{NN}}$). At the initial proper time, $\frac{\partial <p>_{i}}{\partial <\epsilon>_{i}}=\frac{<p>_{i}}{<\epsilon>_{i}}=c_{s}^{2}$ in QPM EOS~\cite{pks1} and thus $<p>_{i}=c_{s}^{2}<\epsilon>_{i}$.

\subsection{Constant Pressure Contour and Radius of Screening Region} 
Since the cooling law for pressure cannot be analytically solved for $\tau$ and, therefore, we use a trick to determine the radius of screening region. Writing the cooling law of pressure as follows :

\begin{equation}
p(\tau,r)=A+\frac{B}{\tau^{q}}+\frac{C}{\tau} +\frac{D}{\tau^{c_s^2}},
\end{equation}
where $A$, $B$, $C$ and $D$ are constants related to $c_1$, $c_2$ and $c_3$ as : $A=-c_1$, $B=c_2 c_s^2$, $C=\frac{4\eta q}{3(c_s^2-1)}$ and $D=c_3$. Writing the above equation at the initial time $\tau=\tau_i$ and at screening time $\tau=\tau_s$ we get : 
\begin{equation}
p(\tau_i,r)=A+\frac{B}{\tau_i^{q}}+\frac{C}{\tau_i}+\frac{D}{\tau_i^{c_s^2}}=p(\tau_{i},0)h(r),
\end{equation} 
and
\begin{equation}
p(\tau_s,r)=A+\frac{B}{\tau_s^{q}}+\frac{C}{\tau_s}+\frac{D}{\tau_s^{c_s^2}}=~p_{QGP}.
\end{equation}
Here $p_{QGP}$ is the QGP pressure as determined by EOS in QPM~\cite{pks,pks1}. Solving Eqs. (11) and (12) numerically and equating the screening time $\tau_s$ to the dilated formation time of quarkonia $t_{F}$ (=$\gamma \tau_F$ where $\gamma=E_{T}/M_{\psi}$ is the Lorentz factor associated with the transverse motion of the $c-\bar{c}$ pair, $M_{\psi}=3.1$ GeV and $\tau_{F}$ is the proper time required for $c-\bar{c}$ pair in the formation of $J/\psi$), we can find the radius of the screening region $r_{s}$. The screening region is defined as a region where temperature is more than the dissociation temperature so that the quarkonia formation becomes unlikely inside that region~\cite{mmish}. Hence the pair will in all probability escape the screening region and form quarkonia only if $|\vec r_{\psi}+\vec v t_F|\ge r_s$ where $\vec r_{\psi}$ is the position vector at which the charm-quark pair is created.

The above kinematic condition takes a simplified form by assuming that $J/\psi$ is moving with transverse momentum $p_T$. Thus the above escape condition can be expressed as a trigonometric condition~\cite{mmish} :

\begin{equation}
\cos \phi\ge Y ;\quad  Y=\frac{(r_s^2-r_{\psi}^2)m-\tau_F^2p_T^2/m}{2r_{\psi}\tau_F p_T},
\end{equation}
where $\phi$ is the angle between the transverse momentum ($p_{T}$) and the position vector $\vec r_{\psi}$ and $r_{\psi}=|\vec r_{\psi}|$ with $m=M_{\psi}$.
\subsection{Survival Probability}
Assuming the radial probability distribution for the production of $c\bar c$ pair in hard collisions at transverse distance $r$ as 
\begin{equation}
f(r)\propto\left(1-\frac{r^2}{R_T^2}\right)^{\alpha}\theta(R_T-r).
\end{equation}    
Here we take $\alpha=1/2$ in our calculation as used in Ref.~\cite{chu}. Then, in the colour screening scenario, the survival probability for the quarkonia can easily be obtained as~\cite{mmish,chu} :

\begin{equation}
S(p_T,N_{part})=\frac{2(\alpha+1)}{\pi R_T^2}\int_0^{R_T}dr r \phi_{max}(r)\left\{1-\frac{r^2}{R_T^2}\right\}^{\alpha},
\end{equation}
where the maximum positive angle $\phi_{max}$ allowed by Eq. (13) becomes~\cite{mmish} :
$$
\phi_{max}(r)=\left\{\begin{array}{rl}
\pi     & \mbox{~~if $Y\le -1$}\\
\pi-\cos^{-1}|Y|  & \mbox{~~if $0\ge Y\ge -1$}\\
\cos^{-1}|Y| & \mbox{~~$0\le Y\le -1$}\\
 0        & \mbox{~~$Y \ge 1$} 
\end{array}\right.
$$
Generally experimentalists measure the quantity namely $p_T$ integrated nuclear modification factor and, therefore, the theoretical $p_T$ integrated survival probability is of paramount importance in comparing with the experimental results. It is defined as follows :

\begin{equation}
S (N_{part})=\frac{\int_{p_{Tmin}}^{p_{Tmax}}S(p_T,N_{part})dp_T}{\int_{p_{Tmin}}^{p_{Tmax}}d p_T}.
\end{equation} 
    
It has been found that only about $60\%$ of the observed $J/\psi$ come from hard collisions whereas $30\%$ is from the decay of $\chi_c$ and remaining $10\%$ from the $\psi^{'}$~\cite{mmish}. Therefore, the net survival probability of $J/\psi$ in the presence of QGP medium is :

\begin{equation}
S_{j/\psi} =0.60\langle S_{J/\psi}\rangle_{p_T}+0.30 \langle S_{\chi_c}\rangle_{p_T}+0.10 \langle S_{\psi^{'}}\rangle_{p_T}
\end{equation}
\begin{table}
\caption{Values of the parameters.}
\begin{tabular}{l|l|l|l|l|l}
\hline\hline
~  & $T_i(GeV)$ & $p_i(GeV^4)$ & $s_i(GeV^3)$ & $\alpha$ & $\beta$\\\hline
SPS & 0.5 & 0.25 & 2.009 & 0.5 & 1.0 \\\hline
RHIC & 0.5 & 0.25 & 2.009 & 0.5 & 1.0\\\hline
LHC & 1.0 & 4.5 & 16.41 & 0.5 & 1.0\\
\hline\hline
\end{tabular}
\end{table}

\begin{table}
\caption{Masses, formation times and dissociation temperatures of the quarkonia.}
\begin{tabular}{l|l|l|l|l}
\hline\hline
    ~    & $m$(GeV) & $\tau_F$(fm) & Set I  ($T_D/T_c$) & Set II ($T_D/T_c$)\\\hline
$J/\psi$ & 3.1 & 0.89 & 2.1 & 2.1 \\\hline
$\chi_C$ & 3.5 & 2.0 & 1.16 & 1.5\\\hline
$\psi^{'}$ & 3.7 & 1.5 & 1.12 & 1.5\\
\hline\hline
\end{tabular}
\end{table}
In our calculation, we use $T_{c}=0.17$ GeV in accordance with the recent lattice QCD results~\cite{mmish}. Similarly initial tempertaure ($T_{i}$), pressure density ($p_{i}$), entropy density ($s_{i}$) at proper time $\tau_{i}$ along with $\alpha$ and $\beta$ at different energies are tabulated in Table. 1. The value of $T_{i}$, $p_{i}$ and $s_{i}$ are taken in accordance with our QPM results~\cite{pks,pks1}. Other parameters like masses ($m$), formation time ($\tau_{F}$) and two different sets of dissociation temperatures ($T_{D}/T_{c}$) labelled as set I and set II for different quarkonia states are given in Table. 2~\cite{mmish,satz}. The reason behind the use of two different sets of dissociation temperature will be discussed later. 


\begin{figure}[!ht]
    \centering
        \includegraphics[height=20em]{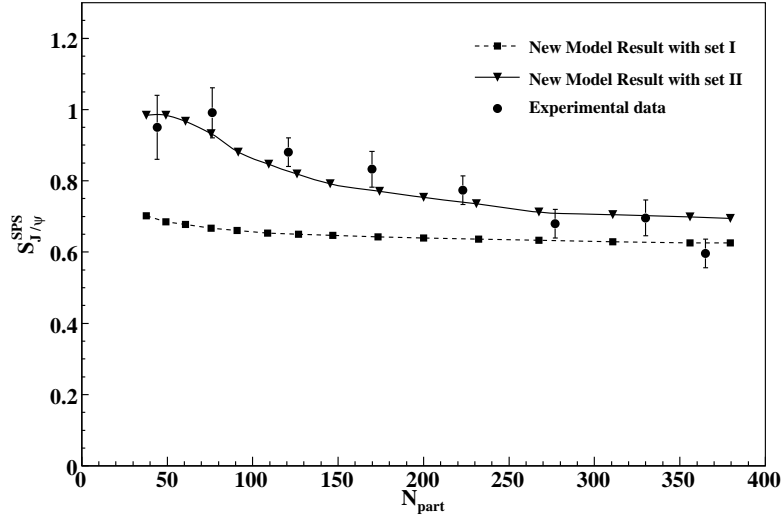}
    \label{fig:Fig 2}
    \caption{Variation of $p_T$ integrated survival probability of $J/\psi$ at SPS energy, ($S_{J/\psi}^{SPS}$) with respect to $N_{part}$ in our present model using two different sets of dissociation temperatures (as given in Table. 2). CNM normalized experimental data points are taken from Ref.~\cite{na60}.}
\end{figure}
\begin{figure}[!ht]
    \centering
        \includegraphics[height=20em]{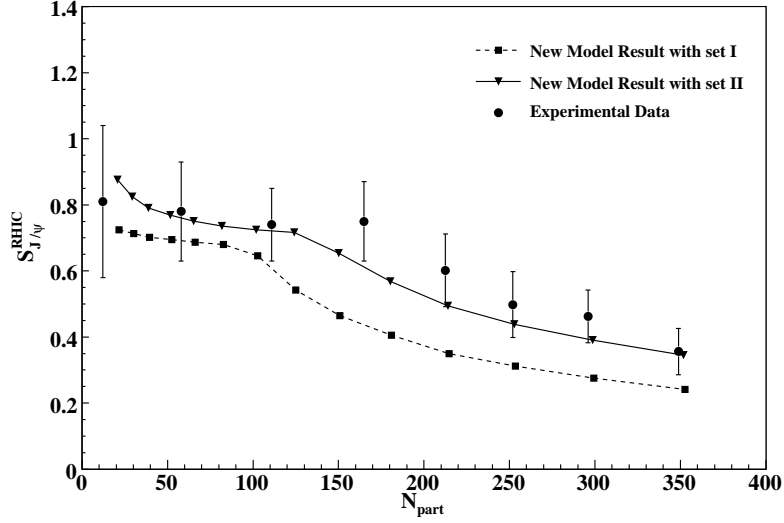}
    \label{fig:Fig 3}
    \caption{Same as in Fig. 1 but at RHIC energy i.e., ($S_{J/\psi}^{RHIC}$) with respect to $N_{part}$. CNM normalized experimental data points are taken from Ref.~\cite{gunji}.}
\end{figure}

\begin{figure}[!ht]
    \centering
        \includegraphics[height=20em]{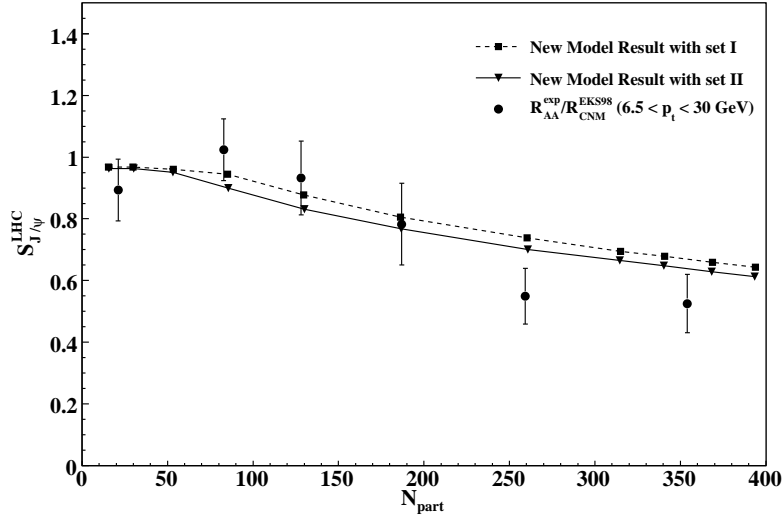}
    \label{fig:Fig 4}
    \caption{Variation of $p_T$ integrated survival probability of $J/\psi$ at LHC energy, ($S_{J/\psi}^{LHC}$) with respect to $N_{part}$. Experimental data points are $R_{AA}^{LHC}$ for prompt $J/\psi$~\cite{hepe} normalized by contribution due to CNM effect~\cite{cnm}.}
\end{figure}

\begin{figure}[!ht]
    \centering
        \includegraphics[height=20em]{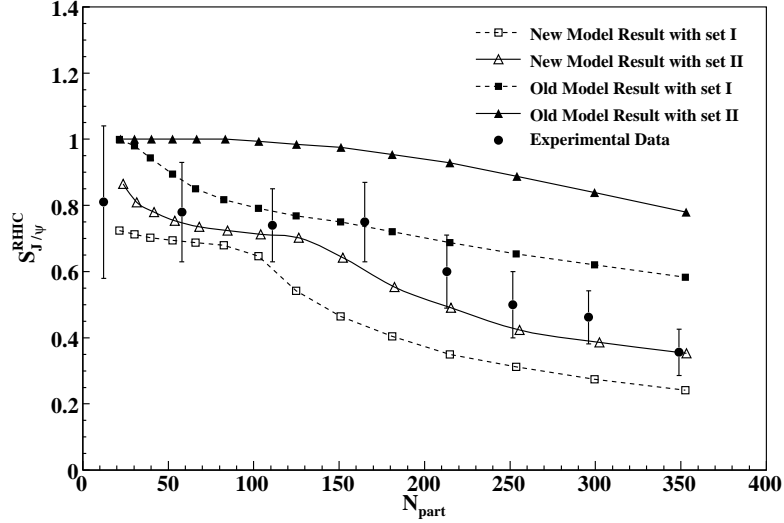}
    \label{fig:Fig 1}
    \caption{Variation of $p_T$ integrated survival probability of $J/\psi$ at RHIC energy, ($S_{J/\psi}^{RHIC}$) with respect to $N_{part}$ in our new model and a comparison with the old model result in which we use bag model as QGP EOS along with ideal hydrodynamics. Experimental data points are taken from Ref.~\cite{gunji}.}
\end{figure}

\begin{figure}[!ht]
    \centering
        \includegraphics[height=20em]{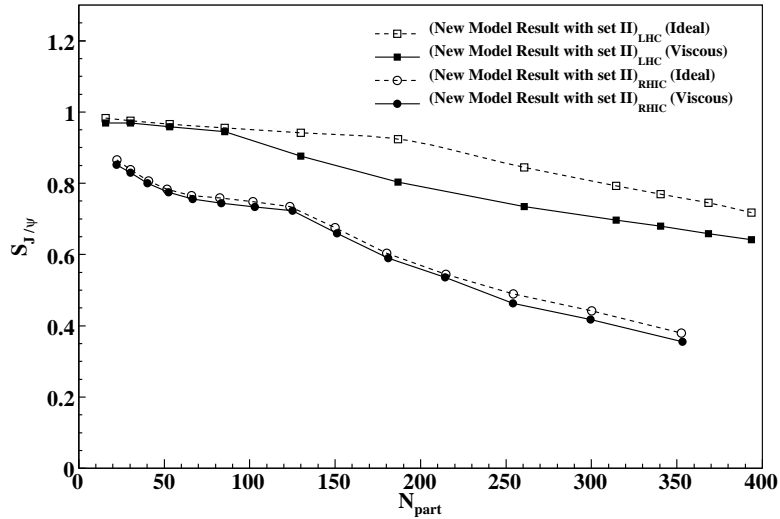}
    \label{fig:Fig 5}
    \caption{Variation of $p_T$ integrated survival probability of $J/\psi$ at RHIC and LHC energies with respect to $N_{part}$ calculated within our new model with dissociation temperature of set II in viscous as well as in ideal hydrodynamics framework.}
\end{figure}


\section{Results and Discussions}

In some recent lattice calculations~\cite{lida,umed}, after constant mode correction, there is no hint for any suppression of $\chi_{c}$ and $\psi^{'}$ instantly after the transition temperature $T_{c}$ as pointed out by different studies~\cite{satz,datt,ohno,aart,jako,oktay}. Moreover, a recent calculation based on QCD sum rules shows that $J/\psi$ quickly dissociates into the continuum and almost disappears completely at temperatures between $1.0~T_{c}$ and $1.1~T_{c}$~\cite{gubler}. Due to above uncertainty regarding the dissociation temperatures of different quarkonia states, we wish to see the effect of $T_{d}$ in simultaneous explaining the data of RHIC, SPS and LHC experiments. Thus we take two different sets of dissociation temperatures for quarkonia states ( labelled as set I and set II as given in Table. 2). Fig. 1 presents the variation of $p_T$ integrated survival probability $S_{J/\psi}^{SPS}$ with $N_{part}$ obtained from our present model for the two sets (labelled as set I and set II ) and we have shown comparison with the CNM normalized experimental data~\cite{gunji}. Moreover, one should keep in mind that from here onwards, we use the same $p_{T}$ range in all our calculations as used in the related experimental data. We observe that our new model with set II yields a reasonable agreement with the experimental data in comparison to the results obtained by the model with set I. Furthermore, we find that the changes in $T_{d}$ values produce more  impact on the peripheral collisions rather than on central collisions.

 Fig. 2 demonstrates the variation of $S_{J/\psi}^{RHIC}$ with respect to $N_{part}$ as obtained from our model with set I and set II, and compare our results with the experimental data again normalized with CNM effect~\cite{gunji}. We again observe that the model with dissociation temperatures given by set II compare well with the data in comparison to the model with set I. At RHIC energy the difference in the results of two models is significant because when we increase the dissociation temperatures for $\chi_{c}$ and $\psi{'}$, the energy deposited in the collision at RHIC energy is still not sufficient to melt all the $\chi_{c}$s and $\psi{'}$s even in the most central collisions and thus they still provide a significant contribution to the survived $J/\psi$s. 

 Fig. 3 shows the variation of $S_{J/\psi}^{LHC}$ with respect to $N_{part}$ obtained in our model with set I and II at LHC energy. Here we find that there is almost no difference in the survival probability calculated with set I or II for all collision centralities. We get lesser suppression in $J/\psi$ at LHC in comparison to RHIC. However, one should keep in mind the integrated $p_{T}$ range here. For LHC, we include quarkonia having $p_{T}$ upto $30$ GeV. We also present the recent experimental data of nuclear modification factor ($R_{AA}$) of prompt $J/\psi$ at mid-rapidity ($|y|~<~2.4$) obtained by CMS collaboration~\cite{hepe} and normalized by cold nuclear matter (CNM) contribution at LHC~\cite{cnm}. The quantitative as well as qualitative agreement of our model results with the CNM normalized experimental data appears quite good. 

Fig. 4 shows the variations of $p_T$ integrated survival probability of $J/\psi$ at RHIC energy (i.e. $\sqrt{s_{NN}}=200 GeV$) with respect to $N_{part}$. Old model represents the results where a bag model EOS for QGP along with ideal hydrodynamics~\cite{mmish} is employed to calculate the $p_T$ integrated survival probability with two sets (i.e., set I and set II) of dissociation temperatures. Blank square and blank triangular symbols represent the results obtained in our present model with quasi-particle model EOS for QGP medium together with the tool of viscous hydrodynamics. The comparison shows a significant effect of the EOS of QGP used in the calculation of survival probability for $J/\psi$. We have also plotted the experimental data~\cite{gunji} (as shown by solid circles in Fig. 2) normalized with CNM effect~\cite{gunji}. We find the present model results employing QPM EOS for QGP along with viscous hydrodynamics give a reasonably good agreement with the experimental data. Again the present model with set II describes the data more suitably in comparison to all other calculations. This show the importance of QPM EOS for QGP over bag model. Furthermore, in the bag model equation of state~\cite{mmish} we have used an additional arbitrary parameter $\xi$ in the Bjorken's formula for initial energy density. Its value $(\xi=5)$ was chosen in such a way that the resulting initial energy density becomes equal to what has been predicted by the self-screened parton cascade model~\cite{esko}. However, in the present model using QPM as EOS, the energy density does not require any such parameter.

 In Fig. 5, we present our new model results for  $S_{J/\psi}^{LHC}$ as well as $S_{J/\psi}^{RHIC}$ with $N_{part}$ and also show a comparison with results obtained in our present model when we put shear viscosity ($\eta$) equal to zero (i.e., considering ideal hydrodynamic framework). We find that the results in the two cases (viscous and ideal) almost overlap on each other for the case of peripheral collisions. However, $p_T$ integrated suvival probability as calculated from ideal hydrodynamics yields quite significant difference at larger $N_{part}$ in comparison to the results obtained from viscous hydrodynamics. This difference becomes more significant at higher energy i.e., at LHC in comparison to RHIC energy. Thus, our study suggests that inclusion of viscous effect in the hydrodynamic evolution is also essential for a complete description of $J/\psi$ suppression.

In summary, we have presented here a modified colour-screening model for $J/\psi$ suppression in QGP medium where quasi-particle model (QPM) equation of state for QGP and feed down from higher resonance states (namely, $\chi_c$, and $\psi^{'}$), also dilated formation time for quarkonia and viscous effects of the QGP medium have been properly taken into account. Furthermore, we assume that the QGP is expanding under Bjorken's boost invariant longitudinal expansion. The above model is thus used to analyze the centrality dependence of the $J/\psi$ suppression data from SPS, RHIC and LHC experiments. We notice that the centrality dependent and $p_T$ integrated survival probability as obtained in the current model accounts well the experimental data in a consistent manner at all available energies. We have also compared our present model with the previous model results~\cite{mmish} where bag model EOS of QGP has been used. We notice that present model describes the experimental data better in comparison to earlier model results~\cite{mmish}. Thus we conclude that our model using colour screening picture describes all the features of $J/\psi$ suppression data simultaneously at SPS, RHIC and LHC energies with the same set (i.e., set II) of dissociation temperatures for different quarkonia states. Further extension of this work is underway to explain the complete rapidity dependence of the data for charmonium and bottomonium resonances in order to draw a firm conclusion regarding the overall level of suppression arising from QGP picture alone.  
\noindent
\section{Acknowledgments}
One of the author (M. Mishra) is grateful to the Department of Science and Technology (DST), New Delhi for financial assistance as FAST-Track Young Scientist project. M. Mishra also acknowledges the financial support from Summer Research Fellowship programme, of Indian Academy of Sciences, Bangalore. P. K. Srivastava thanks the University Grant Commission, New Delhi for financial support.
\pagebreak 

\section*{\LARGE Appendix A}
\subsection*{\underline{(1+1)-Dimensional Cooling Law for Temperature}}

\begin{figure}[!ht]
   \centering
       \includegraphics[height=20em]{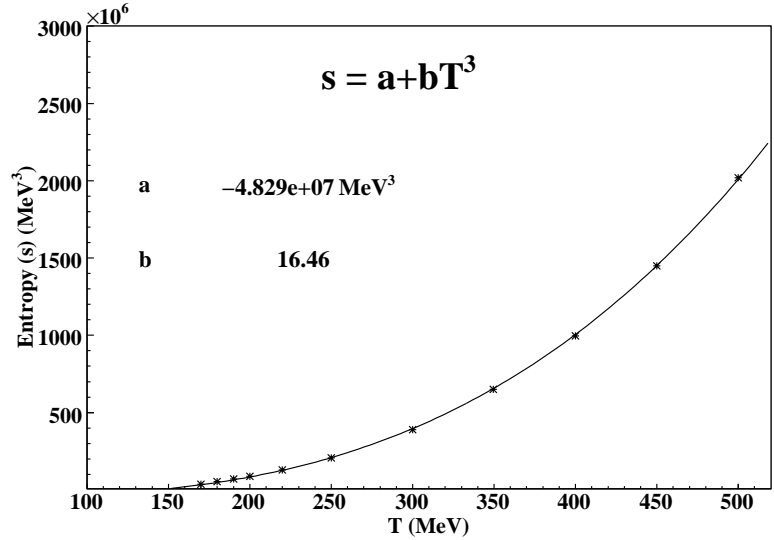}
   \label{fig:Fig 6}
   \caption{Variation of entropy ($s$) with respect to temperature ($T$) calculated using QPM. Solid curve shows a fitting function with $a$ and $b$ as two parameters.}
\end{figure}

In order to determine the explicit proper time ($\tau$) dependence of temperature ($T$), we first start with the entropy equation for (1+1)-dimensional expansion in (3+1)-dimensional space given as follows [{\bf PRD 41}, 2903 (1990)]:

\[\left[\frac{\partial (s\tau)}{\partial \tau}\right]\frac{1}{\tau}=\frac{s}{R\tau},\eqno (A.1)\]

where $s$ is the entropy density of the system and $R$ is the Reynold's number defined as :
\[
R^{-1} = \frac{4/3~\eta+\zeta}{T~s~\tau}\nonumber=\frac{4}{3}\frac{\eta}{T~s~\tau}~ (if~\zeta=0)
\eqno (A.2)\]
Here $\eta$ and $\zeta$ are ahear and bulk viscosities of the medium. Eq. (A.1) can be rewritten as :

\[\left[\frac{\partial s}{\partial \tau}\tau + s\right]\frac{1}{\tau}=\frac{s}{R\tau}\eqno (A.3)\]

We now obtain the entropy curve using QPM as EOS for QGP at zero baryon chemical potential (see Fig. 6). We then derive an empirical relation for entropy density as $s=a+b*T^{3}$ which fits the curve shown above. The values of $a$ and $b$ as obtained from the fit are $-4.829\times 10^{7}$ MeV$^3$ and $16.46$, respectively. Now substituting $s$ in Eq. (A.3) and  treating $R$ as almost independent of $\tau$, we get the following relation :

\[3~b~T^{2}\frac{\partial T}{\partial \tau}=\frac{s}{\tau}\left(\frac{1}{R}-1\right)\eqno (A.4)\]

which gives :

\[\left(\frac{3~b~T^{2}}{a+bT^{3}}\right)\partial T = \left(\frac{1}{R}-1\right)\frac{\partial \tau}{\tau}.\eqno (A.5)\]

We use $u=a+bT^{3}$ and $\partial u= 3bT^{2}\partial T$ and integration of the resulting equation yields 

\[u=c~\tau^{(\frac{1}{R}-1)}=a+bT^{3}.\eqno (A.6)\]

Using the initial time boundary condition $T=T_{0}$ at $\tau=\tau_{0}$, we can determine the constant $c$ as :

\[c=\frac{a+bT_{0}^{3}}{\tau_{0}^{(\frac{1}{R}-1)}}\eqno (A.7)\]

Thus, the cooling law for temperature can finally be written as follows :

\[\frac{T}{T_{0}}=\left(\frac{\tau}{\tau_{0}}\right)^{(\frac{1}{R}-1)}\left[1+\frac{a}{bT_{0}^{3}}\right]-\frac{a}{bT_{0}^{3}}\eqno (A.8)\]

\subsection*{\underline {(1+1)-Dimensional Cooling Law for Energy Density}}
The rate equation for energy dansity in Bjorken's longitudinal expansion scenario for viscous case is : 

\[\frac{\partial\epsilon}{\partial\tau}=-\frac{\epsilon+p}{\tau}+\frac{4\eta}{3\tau^2}.\eqno (A.9)\]

Using the relation : 

\[\epsilon = T\frac{\partial p}{\partial T} - p,\eqno (A.10)\]

and differentiating Eq. (A.10) w.r.t. $\tau$, we get the following relation :

\[\frac{\partial\epsilon}{\partial\tau}=\frac{\partial T}{\partial\tau}\frac{\partial P}{\partial T}+ T\frac{\partial^{2}P}{\partial \tau\partial T}-\frac{\partial p}{\partial\tau}\eqno (A.11)\]

or, in other words,

\[\frac{\partial\epsilon}{\partial\tau}= \left(T\frac{\partial^{2}P}{\partial T^{2}}+\frac{\partial P}{\partial T}\right)\frac{\partial T}{\partial \tau}-\frac{\partial p}{\partial\tau}\eqno (A.12)\]

 Equating Eq. (A.9) with Eq. (A.12) and using $\epsilon+p=T\frac{\partial p}{\partial T}$, we get :

\[-\frac{T}{\tau} \frac{\partial p}{\partial T}+\frac{4\eta}{3\tau^2}=(T\frac{\partial^{2} p}{\partial T^2}+\frac{\partial p}{\partial T}) \frac{\partial T}{\partial\tau} - \frac{\partial p}{\partial\tau}\eqno (A.13)\]

Writing Eq. (A.9) as :

\[\frac{\partial\epsilon}{\partial\tau}=-\frac{T}{\tau} \frac{\partial p}{\partial T}+\frac{4\eta}{3\tau^2}\eqno (A.14)\]

Differentiating the above Eq. (A.14) with respect to $\tau$ we get 

\[\frac{\partial^{2}\epsilon}{\partial \tau^{2}}=-\frac{\partial}{\partial\tau}\left(\frac{T}{\tau}\right)\frac{\partial p}{\partial T}-\frac{T}{\tau}\frac{\partial^{2}p}{\partial\tau\partial T}+\frac{4\eta}{3}\frac{\partial}{\partial\tau}\frac{1}{\tau^{2}}\eqno (A.15)\]

Or, after rearranging terms we get :

\[\tau\frac{\partial^{2}\epsilon}{\partial\tau^{2}}=-\left[\frac{\partial T}{\partial\tau}-\frac{T}{\tau}\right]\frac{\partial p}{\partial T}-T\frac{\partial^{2}}{\partial\tau\partial T}-\frac{8}{3}\frac{\eta}{\tau^{2}}\eqno (A.16)\]

Finally Eq. (A.16) can be rewritten as :

\[\tau\frac{\partial^{2}\epsilon}{\partial\tau^{2}}=-\left(\frac{\partial p}{\partial T}+T\frac{\partial^{2}p}{\partial T^{2}}\right)\frac{\partial T}{\partial\tau}+\frac{T}{\tau}\frac{\partial p}{\partial T}-\frac{8}{3}\frac{\eta}{\tau^{2}}\eqno (A.17)\]

Eliminating $(T\frac{\partial^{2} p}{\partial T^2}+\frac{\partial p}{\partial T}) \frac{\partial T}{\partial\tau}$ with the help of Eq. (A.13), we get

\[\tau\frac{\partial^{2}\epsilon}{\partial\tau^{2}}=-\left[-\frac{T}{\tau}\frac{\partial p}{\partial T}+\frac{4\eta}{3\tau^{2}}+\frac{\partial p}{\partial\tau}\right]+\frac{T}{\tau}\frac{\partial p}{\partial T}-\frac{8}{3}\frac{\eta}{\tau^{2}}\eqno (A.18)\]

Using Eq. (A.14) in Eq. (A.18), we get :

\[\tau\frac{\partial^{2}\epsilon}{\partial\tau^{2}}=-2\frac{\partial \epsilon}{\partial\tau}-\frac{4\eta}{3\tau^{2}}-\frac{\partial p}{\partial\tau}\eqno (A.19)\]

We can rewrite the above Eq. (A.19) as follows:

\[\tau \frac{\partial^{2}\epsilon}{\partial\tau^2} + 2\frac{\partial\epsilon}{\partial\tau} + \frac{4\eta}{3\tau^2}+\frac{\partial p}{\partial\tau}=0.\eqno (A.20)\]

or,

\[\tau \frac{\partial^{2}\epsilon}{\partial\tau^2} + 2\frac{\partial\epsilon}{\partial\tau} + \frac{4\eta}{3\tau^2}+\frac{\partial p}{\partial\epsilon}\frac{\partial\epsilon}{\partial\tau}=0.\eqno (A.21)\]

Now using $\partial p/\partial \epsilon=c_{s}^{2}$ from QPM, we get :
\[\tau^2 \frac{\partial^{2}\epsilon}{\partial\tau^2} + (2+c_s^2)\tau\frac{\partial\epsilon}{\partial\tau}=-\frac{4\eta}{3\tau}.\eqno (A.22)\]
Substituting $\tau=e^{z}$, and $\tau^{2}\frac{\partial^{2}\epsilon}{\partial\tau^{2}}=D(D-1)\epsilon$ where $D=\partial/\partial z$, we get :
\[(D^{2}-D)\epsilon + (c_{s}^{2}+2)D\epsilon = -\frac{4\eta}{3e^{z}}\eqno (A.23)\]
or
\[[D^{2}+ (c_{s}^{2}+1)D]\epsilon = -\frac{4\eta}{3e^{z}}\eqno (A.24)\]
Auxiliary equation for the above differential Eq. (A.24) as
\[m^{2}+ (c_{s}^{2}+1)m = 0\eqno (A.25)\]
The roots of auxiliary equation is $0$ and $-(c_{s}^{2}+1)$. Thus the final cooling law for energy density from Eq. (A.24) can be given as follows :
\[\epsilon=c_1+c_2\tau^{-(c_s^2+1)}+\frac{4\eta}{3c_s^2}\frac{1}{\tau}\eqno (A.26)\]
\subsection*{\underline{(1+1)-Dimensional Cooling Law for Pressure}}
Again starting from energy density rate equation :
\[\frac{\partial\epsilon}{\partial \tau}=-\frac{\epsilon+p}{\tau}+\frac{4\eta}{3\tau^{2}}.\eqno (A.27)\]
It can be rewritten as :
\[\frac{\partial\epsilon}{\partial p}\frac{\partial p}{\partial \tau}=-\frac{\epsilon}{\tau}-\frac{p}{\tau}+\frac{4\eta}{3\tau^{2}}.\eqno (A.28)\]
Using $\partial p/\partial\epsilon=c_{s}^{2}$ from QPM in Eq. (A.28), we get:
\[\frac{\partial p}{\partial \tau}+c_{s}^{2}\frac{p}{\tau}=-c_{s}^{2}\frac{\epsilon}{\tau}+c_{s}^{2}\frac{4\eta}{3\tau^{2}}.\eqno (A.29)\]
Now, using Eq. (A.26), we get the final differential equation for the cooling law of pressure :
\[\frac{\partial p}{\partial \tau}+c_{s}^{2}\frac{p}{\tau}+\frac{c_{s}^{2}}{\tau}c_{1}+\frac{c_{s}^{2}}{\tau}c_{2}\tau^{-(c_{s}^{2}+1)}-\frac{4\eta}{3\tau^{2}}(1+c_{s}^{2})=0.\eqno (A.30)\]
To solve this above partial differential equation, we need to determine its integrating factor (I.F.) which comes out to be $\tau^{c_{s}^{2}}$. Finally the solution of Eq. (A.30) after using this I.F. is:
\[p=-c_1+c_2\frac{c_s^2}{\tau^q}+\frac{4\eta}{3\tau}\left(\frac{q}{c_s^2-1}\right)+c_3\tau^{-c_s^2},\eqno (A.31)\]
where $q=c_{s}^{2}+1$.
\subsection*{Evaluation of constants c1, c2 and c3}
Using the initial time boundary condition $\epsilon=\epsilon_{0}$ at $\tau=\tau_{0}$ in Eq. (A.26), we get the first constraint relation as follows :
\[\epsilon_0=c_1+c_2\tau_0^{-q}+\frac{4\eta}{3c_s^2}\frac{1}{\tau_0}\eqno (A.32)\]
Second boundary condition $\epsilon=0$ at $\tau=\tau^{'}$ (corresponds to temperature $T^{'}=160 MeV$) in Eq. (A.26), yields a second constraint as follows :
\[0 = c_1+c_2\tau^{'-q}+\frac{4\eta}{3c_s^2}\frac{1}{\tau^{'}}\eqno (A.33)\]
Subtracting Eq. (A.33) from Eq. (A.32), we get $c_{2}$ as :
\[c_2=\frac{\epsilon_0-\frac{4\eta}{3c_s^2}\left(\frac{1}{\tau_0}-\frac{1}{\tau^{'}}\right)}{\tau_0^{-q}-\tau^{'-q}},\eqno (A.34)\]
where $\tau^{'}$ is calculated from Eq. (A.8). Similarly $c_{1}$ is obtained from the following relation :
\[c_1=-c_2\tau '^{-q}-\frac{4\eta}{3c_s^2\tau^{'}}.\eqno (A.35)\]
Finally, using initial condition $p=p_{0}$ at $\tau=\tau_{0}$ in Eq. (A.31), we can get the following expression for $c_{3}$ :
\[c_3=(p_0+c_1)\tau_0^{c_s^2}-c_2c_s^2\tau_0^{-1}-\frac{4\eta}{3}\left(\frac{q}{c_s^2-1}\right)\tau_0^{(c_s^2-1)}.\eqno (A.36)\]

\pagebreak

\newpage

\end{document}